\documentclass[aps,prl,showpacs,twocolumn,floats,epsfig]{revtex4}
\usepackage{amssymb}
\usepackage{amsbsy}
\usepackage{amsmath}
\usepackage{epsfig}
\usepackage{amsfonts} % collection is a set of miscellaneous TeX fonts
\usepackage{bm}% bold math
\usepackage{graphicx} % for improved inclusion of graphics
\usepackage[breaklinks,colorlinks = true,linkcolor = blue,urlcolor  = blue,citecolor = blue,anchorcolor = green,bookmarks=true]{hyperref}
\usepackage{mathrsfs}%\mathscr
\allowdisplaybreaks
\begin{document}

\title {Non-equilibrium dynamics of Bosonic Mott insulators in an electric field}

\author{M. Kolodrubetz$^{1}$, D. Pekker$^{2}$, B. K. Clark$^{1,3}$, and K. Sengupta$^{4}$}

\affiliation{
$^{1}$Department of Physics, Princeton University, Princeton, NJ 08544, USA. \\
$^{2}$Department of Physics, California Institute of Technology, Pasadena, CA 91125, USA. \\
$^{3}$Princeton Center for Theoretical Science, Princeton University, Princeton, NJ 08544, USA.\\
$^{4}$ Theoretical Physics Department, Indian Association for the
Cultivation of Science, Jadavpur, Kolkata-700032, India.}

\date{\today}

\begin{abstract}

We study the non-equilibrium dynamics of one-dimensional Mott
insulating bosons in the presence of a tunable effective electric
field ${\mathcal E}$ which takes the system across a quantum
critical point (QCP) separating a disordered and a translation
symmetry broken ordered phase. We provide an exact numerical
computation of the residual energy $Q$, the log-fidelity $F$, the excess defect 
density $D$, and the order parameter correlation function for a 
linear-in-time variation of ${\mathcal E}$ with a rate $v$. We discuss the temporal and
spatial variation of these quantities for a range of $v$ and for
finite system sizes as relevant to realistic experimental setups [J.
Simon {\it et al.}, Nature {\bf 472}, 307 (2011)]. We show that in finite-sized systems $Q$,
$F$, and $D$ obey Kibble-Zurek scaling, and suggest further
experiments within this setup to test our theory.

\end{abstract}


\maketitle

Ultracold atoms in optical lattices provide us with a unique
opportunity to study both equilibrium phases and non-equilibrium
quantum dynamics of strongly coupled bosonic systems near a quantum
phase transition (QPT)~\cite{bloch1}. One system, which has
been the subject of a recent experimental study, consists of
one-dimensional (1D) Mott insulating (MI) bosons in the presence of
an effective electric field ${\mathcal E}$~\cite{greiner1}. It has
been shown that this system can be described in
terms of an effective quantum dipole model or, equivalently, an
Ising spin model with infinitely strong nearest neighbor coupling in
the presence of both a transverse and a longitudinal
field~\cite{subir1}. It has also been demonstrated in
Ref.~\cite{subir1}, that tuning ${\mathcal E}$ to a critical value
${\mathcal E}_c$ leads to a QPT. In the dipole language, this
transition consists of a change in the ground state from a
dipole vacuum (which corresponds to $\bar n$ bosons at each site) to
one with maximal dipoles (which corresponds to alternating ${\bar
n}-1$ and ${\bar n}+1$ bosons per site). In the spin language, this transition 
is from the paramagnet (PM) to an Ising antiferromagnet (AFM). The 
intermediate QCP belongs to the Ising universality class~\cite{subir1}. 
The appearance of this AFM order has recently been observed using 
a quantum gas microscope~\cite{greiner1}. Theoretical studies of
the phases of the bosonic Mott insulator in an electric field
have also been extended to several 2D lattices~\cite{subir2}.

\begin{figure}
\includegraphics[width=\linewidth]{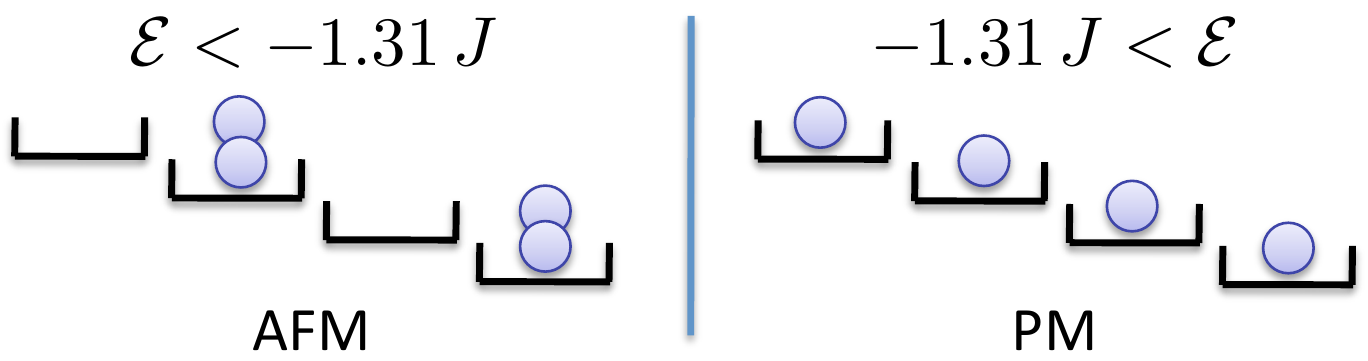}
\caption{(color online) A pictorial representation of the dipole
(AFM) and the vacuum (PM) ground states for MI bosons with ${\bar
n}=1$ across the quantum phase transition. The transition occurs at
$U-{\mathcal E}_c=-1.31$ (for $J=1$).} \label{fig1}
\end{figure}

The study of non-equilibrium dynamics of closed quantum systems have
seen tremendous progress in recent years~\cite{rev1}. One reason for this 
intense effort has been the
possibility of experimentally realizing these dynamics using
ultracold atoms. Indeed, experiments probing non-equilibrium phenomena 
with strongly coupled 2D bosonic atoms, well described by the 2D Bose-Hubbard model,
have been carried out recently~\cite{bakr1}. The corresponding theoretical studies show a 
reasonable qualitative match with experiments~\cite{trefzger1}. 
For the case of 1D bosonic MI in the presence of an electric field, 
order parameter dynamics following sudden quenches of the electric field
have also been studied theoretically~\cite{sengupta1}. 
However, the case of finite velocity and, in particular, nearly adiabatic ramps of ${\mathcal E}$ has not been previously explored.

In this letter, we probe the non-equilibrium dynamics of the bosonic
Mott insulators in the presence of a linear-in-time varying electric
field (chemical potential gradient)
\begin{align}
{\mathcal E}(t)= {\mathcal E}_i+({\mathcal E}_f-{\mathcal E}_i) t/\tau, \label{eq:Eevolution}
\end{align}
where ${\mathcal E}_i$ and ${\mathcal E}_f$ are the initial and
final values of the electric field, and $\tau$ is the ramp time. We define the 
ramp rate to be $v=\partial_t {\mathcal E}(t)$. We
look at ramps that start from the PM phase with unit boson occupation
per site and end either in the AFM phase across the QPT or at the
QCP. 
Based on the resonant manifold model of Ref.~\cite{subir1}, we provide 
an exact numerical computation for finite
system sizes ($L$), using Exact Diagonalization (ED, $L \le 16$) and
time dependent Matrix Product States (tMPS, $L \le 48$), of the residual energy $Q$, the log-fidelity $F$,
the number of defects $D$ (a defect, for the ramps we address, is
a doubly occupied or empty lattice site), and the defect correlation
function $C_{ij}(t)$. We note that the experimental setup of Ref.~\cite{greiner1} constitutes systems of $L\le 16$ lattice sites and
measures $D$ as well as $C_{ij}(t)$; hence our theoretical
analysis constitutes a quantitatively exact description of the
dynamics of the experimental system providing a
valuable guideline to future experiments.  Further, for finite sized systems, our
analysis reveals the manifestation of universal Kibble-Zurek-like (KZ) scaling
\cite{kibzurek, AP0, RD}. In the past, KZ scaling
has been shown to work well in integrable systems~\cite{KZscalingI}; 
however, its manifestation in non-integrable systems has not been consistently demonstrated~\cite{KZscalingNI}.  Our work thus provides the first realization of KZ scaling in finite-sized non-integrable systems which can be tested with an existing experimental setup. Furthermore, we investigate the crossover from Landau-Zener (LZ) scaling~\cite{vit1,AP0} to KZ scaling in finite-size systems, and show that it collapses onto a universal functional form, as suggested in Ref.~\cite{AP}.

{\it Model ---\/} Bosonic atoms in a tilted optical lattice 
(i.e. in an effective time dependent electric field) are well described using 
the Bose-Hubbard Hamiltonian \cite{subir1}
\begin{align}
H_B(t) = -t_0 \sum_{\langle ij\rangle} b_i^{\dagger} b_j + \frac{U}{2} \sum_i
n_i (n_i-1) - {\mathcal E}(t) \sum_i i \cdot n_i \nonumber
\end{align}
where $\langle ij\rangle$ denotes the sum over nearest-neighbors, 
$b_i$ the boson annihilation operator at site $i$, and $n_i=b_i^{\dagger}b_i$ the boson density operator. Under the condition $U, {\mathcal E}(t) \gg |U-{\mathcal E}(t)|, t_0$, the low-energy dynamics of $H_B$ are captured by an effective dipole
model~\cite{subir1}
\begin{eqnarray}
H_{d}(t) &=& [U-{\mathcal E}(t)]\sum_{\ell} d_{\ell}^{\dagger}
d_{\ell} - J \sum_{\ell} (d_{\ell}^{\dagger} + d_{\ell}),
\label{ham1}
\end{eqnarray}
where $d_{\ell}= b_i b_j^{\dagger}/\sqrt{n_0(n_0+1)}$ denotes the
dipole annihilation operator which lives on a link $\ell$ between
the neighboring sites $i$ and $j$ as schematically shown in Fig.~\ref{fig1}, 
$n_0$ the boson occupation of parent Mott insulator, and
$J=t_0\sqrt{n_0(n_0+1)}$ the amplitude for creation or
annihilation of a dipole. Henceforth, we shall use the units in
which $\hbar=1$, $J=1$, and restrict ourselves to $n_0=1$. The dipoles satisfy two constraints;
first, there can be only one dipole on any link $\ell$ which renders
$d_{\ell}^{\dagger} d_{\ell} \le 1$, and second, two consecutive
links can not be simultaneously occupied by dipoles
$d_{\ell}^{\dagger} d_{\ell+1}^{\dagger} d_{\ell +1} d_{\ell} =0$.
These constraints render $H_d$ non-integrable; however, they also
lead to a significant reduction in the Hilbert space of $H_d$ which
makes ED and tMPS the methods of choice for
studying the dynamics of the model. In addition, we restrict ourselves to studying 
$H_d$ with periodic boundary conditions so as to approach the thermodynamic 
limit with smaller systems. We note that the dipole model can be represented in terms of an Ising-like spin-model via the
transformation: $S_{\ell}^z=
1/2-d_{\ell}^{\dagger} d_{\ell}$, $S_{\ell}^{x(y)} = (-i)(d_{\ell}
+(-) d_{\ell}^{\dagger})/2$~\cite{subir1,greiner1}. 
Note that $\langle S_{\ell}^z \rangle$ is the order
parameter for the transition from the PM (dipole vacuum) to the AFM
(maximal dipole density) state.

To study the dynamics within exact diagonalization, we evolve the
wave function using the time dependent Hamiltonian of
Eq.~\eqref{ham1}, in which the electric field is tuned linearly in
time according to Eq.~\eqref{eq:Eevolution}
\begin{align}
i \hbar \partial_t |\psi(t)\rangle = H_d(t) |\psi(t)\rangle.
\label{eq:Schrodinger}
\end{align}
We supplement the time evolution with the initial condition $|\psi(t=0)\rangle =
|\psi_G\rangle_i$, where $|\psi_G\rangle_i$ is the ground state wave
function of the initial Hamiltonian. Integrating
Eq.~\eqref{eq:Schrodinger} from $t=0$ to $t=\tau$ we obtain the
wave function at the end of the ramp $|\psi(\tau)\rangle$.

We supplement our exact diagonalization studies by tMPS, which allows
us to study larger system sizes. tMPS represents the wave function as
$|\psi\rangle= \sum_{\{\sigma_i\}} M_1^{\sigma_1} M_2^{\sigma_2} ...                                                                                          
M_L^{\sigma_L} |\sigma_1,\sigma_2 ...\rangle$, where
$M_i^{\sigma_i}$ are a set of $\chi$ by $\chi$ matrices indexed by site $i$ and spin
$\sigma_i$~\cite{tmps}.
We take advantage of the fact that in the reduced Hilbert
space the Hamiltonian $H_d(t)$, Eq.~\eqref{ham1}, is a sum of single site
Hamiltonians to perform time evolution.  We evolve in time via
$|\psi(t+\epsilon)\rangle = P \exp(-i \epsilon H_d(t))$ by first exactly applying the
single site hamiltonian $H_d(t)$  and then projecting out configurations with
neighboring dipoles  using the projection operator $P$.                                                                                            
This projection increases the MPS
bond dimension which is then reduced back to its original value $\chi$ introducing a
small error.                                                                                                               
The method becomes exact in the limit
$\epsilon \to 0$ and matrix size $\chi \to \infty$, and its convergence
has been numerically checked by extrapolating in these parameters~\cite{startPoint}.
Ground states are found by evolving the same Hamiltonian to large imaginary
time at fixed $\mathcal E$.

\begin{figure*}
\includegraphics[width=2\columnwidth]{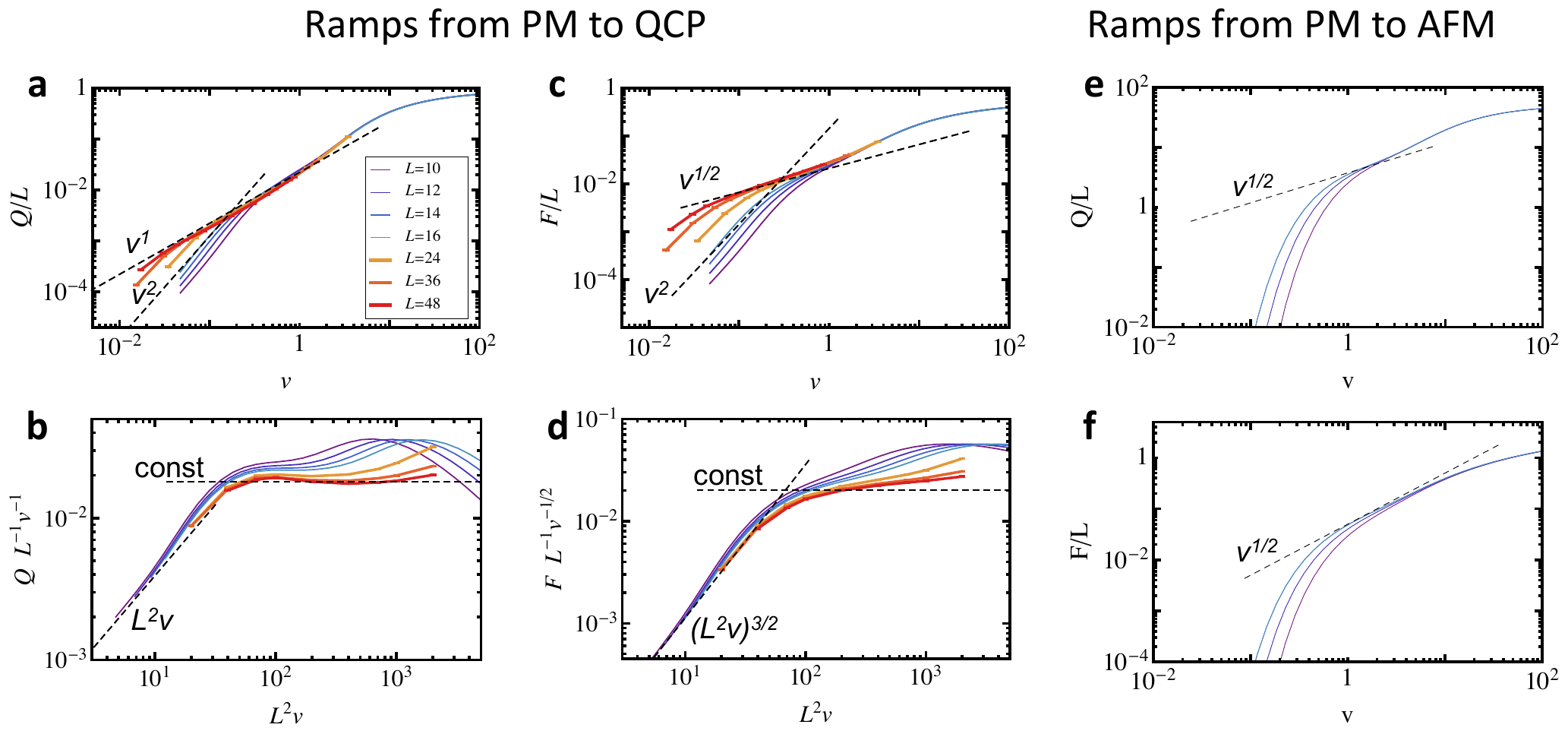}
\caption{(color online) Residual energy $Q$ (a) and log-fidelity $F$ (c) as a function of the ramp rate $v=\partial_t {\mathcal
E}(t)$ for various system sizes from $10\le L \le 16$ (ED) and $24\le L \le 48$ (tMPS, error bars correspond to time step extrapolation error, bond dimension errors are below $10^{-4}$).
The ramps start in the PM ($U-{\mathcal E}_i=100$) and end at the
QCP ($U-{\mathcal E}_f=-1.31$). Dashed lines show the indicated power laws. 
Note the extension of the intermediate Kibble-Zurek-like scaling regime to
lower values of $v$ for larger system sizes. (b) and (d) show
finite size scaling collapse for $Q$ and $F$ for several $L$.  
(e) and (f) show $Q$ and $F$ as a function of $v$ for ramps starting in the PM ($U-{\mathcal E}_i=100$)
and ending in the AFM phase ($U-{\mathcal E}_f=-100$) with $10 \le L \le 14$. Note the change in
$Q$ power law from $v^{1}$ to $v^{1/2}$.}
\label{fig2}
\end{figure*}

For this work, the specific observables of interest will be the
residual energy $Q$, the log-fidelity $F$, the defect number
$n_d$ (i.e. the number of sites with even parity of the boson occupation number), 
excess defect number $D$, and the spin-spin correlation 
function $C_{ij}$ which, at any instant $t$, are given by
\begin{align}
Q(t) &= \langle \psi(t)|H(t)|\psi(t)\rangle - E_G(t),
\label{reseqn} \\
F(t) &= \log [|\langle \psi(t)|\psi_G(t)\rangle|^2],
\label{fideqn} \\
n_d(t) &= \langle \psi(t)|{\sum}_{\ell} \left(1+2 S^z_{\ell}\right)|\psi(t)\rangle, \label{deneqn} \\
D(t)&=n_d(t) -  \langle \psi_G(t)|{\sum}_{\ell} \left(1+2 S^z_{\ell}\right)|\psi_G(t)\rangle,\\
C_{ij}(t) &= \langle \psi(t)|S^z_{i} S^z_{j}|\psi(t)\rangle,
\label{correqn}
\end{align}
where $E_G(t)$ $[|\psi_G(t)\rangle]$ corresponds to the
ground state energy [wave function] of the Hamiltonian $H_d(t)$. For notational
brevity, we shall drop the index $t$ from observables when
evaluating them at the end of the ramp ($t=\tau$).

We begin with a discussion of $Q$ and $F$ for finite size systems
($L \le 48$) undergoing ramps from the PM phase ($U-{\mathcal
E}_i=100$) to the QCP ($U-{\mathcal E}_f=U-{\mathcal E}_c=-1.31$) in
time $\tau$. For a finite size system which is always gapped, $Q$
and $F$ behave differently than their counterparts in the
thermodynamic limit ($L \to \infty$) for which one expects KZ-like
scaling to manifest in both $Q \sim v^{(d+z)\nu/(z\nu+1)}$ and
$F  \sim v^{d\nu/(z\nu+1)}$ for small $v$ (large $\tau$). Here
$d=1$ is the dimensionality, $z=1$ is the dynamical critical
exponent, and $\nu=1$ the correlation length critical exponent.
For very fast ramps, the wave function does not have time to 
evolve during the dynamics, and therefore the behavior of finite-
and infinite-sized systems is similar but not universal.
As the ramp rate becomes slower ($v\sim1$), KZ scaling sets in for both infinite- and 
finite-sized systems. However, while KZ scaling is expected to persist 
to infinitely slow ramps in the thermodynamic limit, for finite-sized systems
it is cut off for ramps slower than a critical ramp rate
$v_c(L) \sim L^{-(1/\nu+z)}$. For $v \le v_c(L)$, $Q$ and $F$
are expected to scale as $v^2$ as in gapped systems
\cite{vit1, AP0}. These expectations may be formalized in the form of
scaling functions
\begin{align}
Q &\sim L^d v^{(d+z)\nu/(z\nu+1)} g_r(v L^{1/\nu+z}), \label{eq:Qscale}\\
F &\sim L^d v^{d\nu/(z\nu+1)} f_r(v L^{1/\nu+z}), \label{eq:Fscale}
\end{align}
where $g_r(x \ll 1) \sim x^{2-(d+z)\nu/(z\nu+1)}$,  $f_r(x \ll 1)
\sim x^{2-d\nu/(z\nu+1)}$ (very slow, i.e. LZ regime) and $g_r(x \gg 1) \sim
\text{const.}$, $f_r(x \gg 1) \sim \text{const.}$ (KZ
regime)~\cite{AP}.

The above-mentioned expectations are corroborated in panels (a) and
(c) of Fig.~\ref{fig2}, which show the behavior of $Q$ and $F$ as a
function of the quench rate $v^{1}$ for several system sizes $10
\le L\le 48$ on a log-log scale. For both $Q$ and $F$ we find
$v^{2}$ scaling for very slow ramps and KZ-like scaling $Q \sim
v^{1}$ and $F \sim v^{1/2}$ for intermediate ramps. Finally,
for fast ramps, we find non power-law behavior followed by a
plateau. We note that KZ-like scaling only occurs in systems of
sufficient size ($L \ge 10$ for $Q$ and $L \ge 16$ for $F$)~\cite{systemSize}. 
The LZ-KZ scaling crossover 
(Eqs.~\eqref{eq:Qscale} and \eqref{eq:Fscale}) is further
corroborated in panels (b) and (d) of Fig.~\ref{fig2}. These plots
demonstrate both the scaling collapse of $Q$ and $F$ for slower
ramps (points to the left) and their deviations from scaling for
faster ramps (points to the right) in the non-universal regime. The
dashed lines indicate the expected form of the
scaling functions $g_r$ and $f_r$ in both the LZ and the KZ
regimes. Note that for the LZ regime, we see collapse even
for relatively small system sizes ($L\gtrsim10$); however, for the
KZ regime collapse only occurs for larger system sizes ($L\gtrsim
16$ for $Q$ and $L\gtrsim 48$ for $F$).

\begin{figure*}
\includegraphics[width=2\columnwidth]{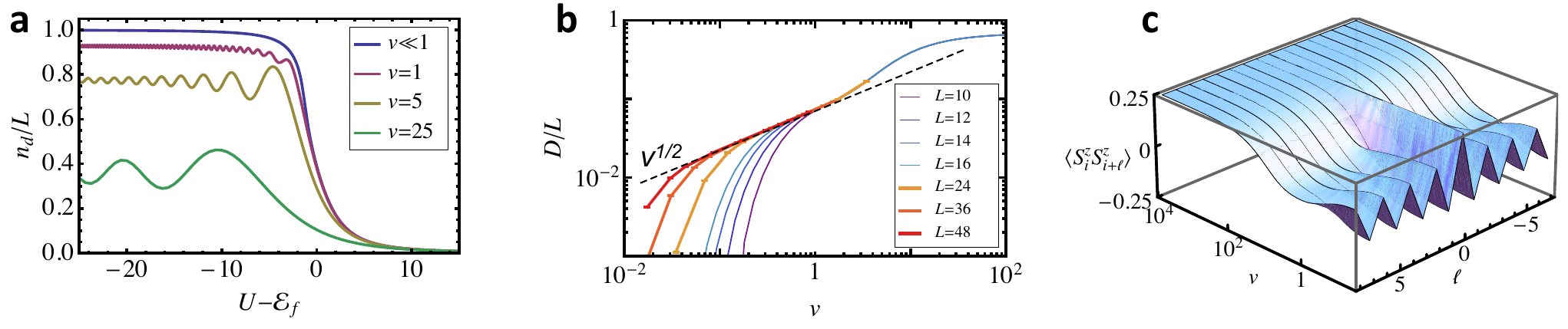}
\caption{(color online) (a) $n_d/L$ as a function of the end point 
of the ramp $U-{\mathcal E}_f$ for several ramp rates (all ramps 
start in the PM phase $U-{\mathcal E}_i=50$). (b) $D$ as a
function of $v$ for several $L$ showing KZ scaling 
($U-{\mathcal E}_i=100$, $U-{\mathcal E}_f=-1.31$).
(c) Spin-spin correlation function $C_\ell=\langle S^z_i S^z_{i+\ell}\rangle$ at the end of
the ramp as a function $\ell$ and the ramp rate $v$ for $L=16$. 
} \label{fig4}
\end{figure*}

Next, we study ramps that cross the QCP. As the system progresses
towards the QCP, the gap closes, and excitations are produced.
At some later time, after passing the QCP, the gap reopens, and no
new excitations can be created. However, although the number of
excitations is conserved in the final adiabatic part of the ramp,
the energies of individual excitations continue changing. Therefore,
in ramps across the QCP, we can expect to see KZ-like scaling for
$F$ but not for $Q$. These expectations are corroborated in panels
(e) and (f) of Fig.~\ref{fig2}, in which we plot $Q$ and $F$ as a
function of the ramp rate $v^{1}$ for ramps from PM ($U-{\mathcal
E}_i=100$) to AFM phase ($U-{\mathcal E}_f=-100$). We find KZ-like
scaling in $F\sim v^{1/2}$ as expected; interestingly $Q$ also
scales as $v^{1/2}$. This happens due to the fact that during
the adiabatic part of the ramp, almost all excitations are converted
into singly occupied sites, with energy per excitation of $\sim |U-
{\mathcal E}_f|$. Therefore $Q$ becomes identical to the number of
excitation and hence scales in the same way as $F$. Thus our
analysis suggest anomalous scaling of $Q$ in ramps across the QCP
should be interpreted with caution.

Having obtained the scaling behavior, we concentrate on its
manifestation for experimentally observable quantities. We note that
the existing experimental setup \cite{bakr1} focuses on imaging the
parity of the number of bosons on each site, after projecting the
wave function into a Fock state~\cite{bloch1, greiner1,bakr1}.
Effectively, this imaging counts the number of empty and doubly occupied
sites (i.e. ``defects" on the PM side). In addition to measuring this
defect number $n_d$, the existing experiments can measure the
defect-defect correlation function $g_d(\ell)=\langle n_d(i) n_d(i+\ell)
\rangle \sim \langle S_i^z S_{i+\ell}^z\rangle= C_{i,i+\ell}$ at the end of the
ramp. In Fig.~\ref{fig4}a, we plot the defect density $n_d$ as a function 
of the $U-{\mathcal E}_F$. We note that the final saturation
value of $n_d$ in the AFM phase is a decreasing function of the ramp
rate and lies between those for a nearly adiabatic ramp and a sudden
quench. In Fig.~\ref{fig4}b, we show that the excess defect number $D$ demonstrates
similar finite-size KZ scaling as $F$. Since $n_d$ is experimentally
measurable, our work demonstrates that finite size scaling can
be observed within an existing experimental setup. Finally, in 
Fig.~\ref{fig4}c, we show the behavior of $C_{i,i+\ell}$ as a function 
of $v$ and $\ell$. We note that for slow ramps
to the QCP one finds an oscillatory behavior of $C_{i,i+\ell}$ indicating the
precursor of the AFM order present in the critical ground state. As
$v$ is increased, the system ceases to reach the final ground
state and these correlations decay. Finally, the state of the system
at the end of fast ramps is essentially the PM ground state, which has no AFM correlations, 
leading to a flat $C_{i,i+\ell}$. These features can be directly picked up in experiments,
which can serve as a test of our theory.

In conclusion, we have investigated universal scaling dynamics of
finite size non-integrable bosonic system following a finite rate
ramp of the effective electric field. Our investigation demonstrates
two scaling regimes (LZ- and KZ-like
scaling) with conventional exponents and thus differ from prior
studies of other non-integrable
systems~\cite{kibzurek,KZscalingI,KZscalingNI} which found various anomalous
scaling exponents. Furthermore, comparing ramps that cross
the QCP (which show anomalous exponents) to those that end at the
QCP (which show expected exponents), we suggest a possible origin of
these anomalous exponents: the adiabatic dynamics of the
excitations following the passage through the quantum critical
regime. Finally, we compute experimentally measurable quantities
such as $n_d$ and $C_{ij}$ and demonstrate that the scaling behavior
studied in this work can be observed in realistic experiments via
measurement of $n_d$.

It is our pleasure to thank A. Polkovnikov for invaluable
discussions. We acknowledge support from the Lee A DuBridge
fellowship (DP), IIAS, PCTS. KS thanks DST, India for support
through grant SR/S2/CMP-001/2009. DP and BKC thank the Aspen 
Center for Physics for its hospitality.


\begin{thebibliography}{99}

\bibitem{bloch1} M. Greiner {\it et al.}, Nature (London) {\bf 415}, 39
(2002).

\bibitem{greiner1} J. Simon {\it et al.}, Nature {\bf 472}, 307
(2011).

\bibitem{subir1} S. Sachdev, K. Sengupta, and S. M. Girvin, \prb
{\bf 66}, 075128 (2002).

\bibitem{subir2} S. Pielawa {\it et al.} Phys. Rev. B {\bf 83},
205135 (2011)

\bibitem{rev1} A. Polkovnikov {\it et al.}, arXiv:1007.5331;
D. Ziarmaga, Adv. Phys. {\bf 59}, 1063 (2010).

\bibitem{bakr1} W. Bakr {\it et al.}, Science {\bf 329}, 547 (2010).

\bibitem{trefzger1} C. Trefzger and K. Sengupta \prl {\bf 106} 095702
(2010).

\bibitem{sengupta1} K. Sengupta, S. Powell, and S. Sachdev, \pra {\bf 69},
053616 (2004).

\bibitem{kibzurek} T. W. B. Kibble, J. Phys. A {\bf 9}, 1387 (1976); W. H. Zurek, Nature {\bf 317}, 505 (1985); 

\bibitem{AP0} A. Polkovnikov and V. Gritsev, Nat. Phys. {\bf 4}, 477 (2008). C. De Grandi and A. Polkovnikov, ''Quantum Quenching, Annealing and Computation," Eds. A. Das, A. Chandra, and B. K. Chakrabati, Lect. Notes in Phys., vol. 802 (Springer, Heidelberg 2010).

\bibitem{RD} M. M. Rams and B. Damski, Phys. Rev. Lett. {\bf 103}. 170501 (2009).

\bibitem{KZscalingI} A. Polkovnikov, Phys. Rev. B {\bf 72}, 161201(R) (2005); R. W. Cherng and L. S. Levitov, Phys. Rev. A {\bf 73}, 043614 (2006); C. De Grandi, V. Gritsev, and A. Polkovnikov, Phys. Rev. B {\bf 81}, 224301 (2010).

\bibitem{KZscalingNI} F. Pollmann, S. Mukerjee, A. G. Green, and J. E. Moore, Phys. Rev. E {\bf 81}, 020101(R) (2010); J.-S. Bernier, G. Roux, C. Kollath, arXiv:1010.5251.

\bibitem{vit1} N. Vitanov and B. M. Garraway, Phys. Rev. A {\bf 53}, 4288 (1996).

\bibitem{AP} C. de Grandi, A. Polkovnikov, and A. Sandvik,
arXiv:1106.XXXX (2011).

\bibitem{tmps} U. Schollwock, Ann. Phys. {\bf 326}, 96 (2011); G.
Vidal, \prl {\bf 91}, 147902 (2003).

\bibitem{startPoint} To speed up tMPS, for slower ramp rates we move the starting point $U-\mathcal{E}_i$ closer to the QCP. We have verified this does not affect the observables.

\bibitem{systemSize} Care must be taken in interpreting data on smaller systems, as these 
appear to show a small segment of anomalous power-law like
scaling in the KZ regime. However, this behavior is
a finite size effect and the universal KZ exponents are restored for larger 
system sizes.  

\end{thebibliography}
\end{document}